\documentclass[11pt]{article}
\usepackage{type1cm,amsmath,hangcaption,graphicx,indentfirst}

\begin{document}


\begin{center} {\tt \Large In press at {\em Foundations of} Physics} \end{center}

\vskip 0.5cm

\begin{center}
{\bf \Large Travelling-Wave Solutions for Korteweg-de Vries-Burgers
Equations through Factorizations}
\end{center}


\bigskip

\begin{center} {\bf O. Cornejo-P\'erez\footnote{\noindent Potosinian Institute of Science and Technology (IPICyT),
Apdo Postal 3-74

$\quad$Tangamanga, 78231 San Luis Potos\'{\i}, S.L.P., Mexico;
e-mail: hcr@ipicyt.edu.mx}, J. Negro\footnote{\noindent Departamento
de F\'{\i}sica Te\'orica, At\'omica y \'Optica, Universidad de
Valladolid, 47071

$\quad$Valladolid, Spain}, L.M. Nieto$^2$ and H.C. Rosu$^1$} \end{center}

\bigskip
\bigskip
\bigskip

\noindent Travelling-wave solutions of the standard and compound
form of Korteweg-de Vries-Burgers equations are found using
factorizations of the corresponding reduced ordinary differential
equations. The procedure leads to solutions of Bernoulli equations
of nonlinearity 3/2 and 2 (Riccati), respectively. Introducing the
initial conditions through an imaginary phase in the travelling
coordinate, we obtain all the solutions previously reported, some of
them being corrected here, and showing, at the same time, the
presence of interesting details of these solitary waves that have
been overlooked before this investigation.

\bigskip




\section*{1. INTRODUCTION}

\noindent In this research we will study nonlinear Korteweg-de
Vries-Burgers (KdVB) equations which play an important role both in
physics and in applied mathematics. First, we will consider the
original KdVB equation \cite{roychoudhury,zwillinger,canosa}, which
can be written as
\begin{equation}
\frac{\partial u}{\partial t} = s\frac{\partial^3 u}{\partial
x^3}-\mu \frac{\partial^2 u}{\partial x^2} -\alpha u\frac{\partial
u}{\partial x}~. \label{L1}
\end{equation}
Second, we will focus on the so-called compound KdVB equation that
has an additional non-linear term \cite{duffy,parkes,wang,xia}
\begin{equation}
\frac{\partial u}{\partial t} = s\frac{\partial^3 u}{\partial
x^3}-\mu \frac{\partial^2 u}{\partial x^2} -\alpha u\frac{\partial
u}{\partial x} -\beta u^2\frac{\partial u}{\partial x}~. \label{L2}
\end{equation}
In both cases the parameters $s,\mu,\alpha,\beta$ are real constants
which take into account different effects, as non-linearity,
viscosity, turbulence, dispersion or dissipation. As it is well
known, these equations have been used  as  mathematical models for
the propagation of waves on elastic tubes \cite{antar}, and in
particular, the KdVB equation has been explicitly derived in the
case of ion-acoustic shock waves in multi-electron temperature
collisional plasma \cite{shukla}. These ion-acoustic shock waves
have been observed in an unmagnetized plasma \cite{nakamura}.
Standard applications are encountered in hydrodynamics and in the
theory of diffusive cosmic-ray shocks \cite{hydro,cosmic-ray}. Other
interesting and more recent applications refer to the dynamics of
high-amplitude picosecond strain pulses in sapphire single crystals
in which case a KdVB model has been shown to fit the observed
behavior at all strains and temperatures under study
\cite{muskens02} and explaining the temporal self-shift of
small-amplitude dark solitons due to self-induced Raman scattering
in optical fibers \cite{Kiv90}.

We are interested in travelling-wave solutions of the Eqs.
(\ref{L1})--(\ref{L2}), that is, solutions of the form
\begin{equation}
u(x,t)= \phi(\xi), \quad \xi=x-vt , \label{L3}
\end{equation}
where $v$ is the velocity of the propagation, whose possible values
will be determined later. By substituting  (\ref{L3}) in (\ref{L2})
and assuming $s\neq 0$, we get an ordinary differential equation
(ODE) to be satisfied by $\phi(\xi)$:
\begin{equation}
\frac{d^3 \phi}{d\xi^3} -\frac{\mu}{s} \frac{d^2 \phi}{d\xi^2} +
\left( \frac{v}{s} -\frac{\alpha}{s} \phi -\frac{\beta}{s}
\phi^2\right) \frac{d \phi}{d\xi} =0. \label{L4}
\end{equation}
A further simplification of this equation is accomplished by means
of the following linear transformation of the dependent and
independent variables
\begin{equation}
\xi=\frac{s}{\mu}\, \theta, \quad \phi(\xi)= \frac{2 \mu^2}{\alpha
s} \, w(\theta), \label{L5}
\end{equation}
which transform Eq.  (\ref{L4}) in
\begin{equation}
\frac{d^3 w}{d \theta^3} - \frac{d^2 w}{d \theta^2} + \left( p -2 w
-3q\, w^2\right) \frac{d w}{d \theta} =0, \label{L6}
\end{equation}
with
\begin{equation}
p=\frac{v s}{\mu^2}, \quad q= \frac{4\beta \mu^2}{ 3 s \alpha^2} .
\label{L7}
\end{equation}
In the sequel, we will use the simplified form (\ref{L6}), which
corresponds to the compound KdVB equation (\ref{L2}) if $q\neq 0$
and to the usual KdVB equation (\ref{L1}) if $q=0$. It is quite
obvious that Eq.  (\ref{L6}) admits a first integral
\begin{equation}
\frac{d^2 w}{d \theta^2} - \frac{d w}{d \theta} + \left( p\, w - w^2
-q\, w^3\right)  = k, \label{L8}
\end{equation}
where $k$ is an arbitrary constant. In the following section we will
summarize the factorization method to find solutions  of general
second order differential equations of the type (\ref{L8}). Then, in
sections 3 and 4, we will apply this method to KdVB equations.


\section*{2. FACTORIZATION OF A SECOND ORDER ODE}

 The factorization method is a well known technique used to find
solutions of ODE. It goes back to some papers of Schr\"odinger in
which he solved some particular examples of the equation bearing his
name \cite{schrodinger}, and it was later developed by Infeld and
Hull \cite{infeldhull} (for more details, see also \cite{oscar} and
the citations quoted therein).

In a more general context, the factorization of non-linear second
order ODE has been previously studied in \cite{berkovich,rosu} for
equations of the form
\begin{equation}
\frac{d^2 {U}}{d \theta^2}- \frac{d {U}}{d \theta}
+F({U})=0,\label{n1}
\end{equation}
where  $F({U})$ is a polynomial function (this is precisely the form
of the two KdVB equations  (\ref{L8}) considered in the previous
section). In these works, Eq. (\ref{n1}) was  factorized as
\begin{equation}
[D-f_{2}({U})][D-f_{1}({U})]{U}(\theta)=0,\label{n2}
\end{equation}
where $D={d}/{d \theta}$. The following grouping of terms has been
proposed in \cite{rosu}
\begin{equation}
\frac{d^2 U}{d \theta^2} -\left(f_{1}+f_{2}+ \frac{df_{1}}{dU} U
\right)\frac{d U}{d \theta} +f_{1}f_{2}U=0,\label{n4}
\end{equation}
and comparing Eq. (\ref{n1}) with Eq. (\ref{n4}), we get the
conditions
\begin{eqnarray}
&&f_{1}(U)\, f_{2}(U)=\frac{F(U)}{U},\label{n5}\\
&&f_{2}(U)+ \frac{d(f_{1}(U)\, U)}{dU}= 1.\label{n6}
\end{eqnarray}
Then, according to  \cite{berkovich}, any factorization like
(\ref{n2}) of an ODE of the form given in Eq. (\ref{n1}), allows to
find a compatible first order ODE
\begin{equation}
[D-f_{1}(U)]U=\frac{d U}{d \theta}-f_{1}(U)U=0,\label{n7}
\end{equation}
whose solution provides a particular solution of (\ref{n1}). In
other words, if by some means we are able to find a couple of
functions $f_{1}(U)$ and $f_{2}(U)$ such that they factorize Eq.
(\ref{n1}) in the form (\ref{n2}), solving the ODE (\ref{n7}) we
will get particular solutions of (\ref{n1}). The advantage of the
factorization presented here is that the two unknown functions
$f_{1}(U)$ and $f_{2}(U)$ can be found easily by factoring the
polynomial expression  (\ref{n5}) in terms of  linear combinations
in rational powers of $U$.

Now, let us apply this technique of finding particular solutions to
the KdVB equations considered in the previous section.

\section*{3. FACTORIZATION OF THE KdVB EQUATION}

 Let us consider first the KdVB equation in the integrated and
simplified form given in (\ref{L8})
\begin{equation}
\frac{d^2 w}{d \theta^2} - \frac{d w}{d \theta} + \left( p \, w -
w^2 \right)  = k. \label{kb1}
\end{equation}
It can be easily proved that this equation can be factorized in the
form (\ref{n2}) by means of a linear combination of powers of $w$
only if $k=0$, which is a very restrictive condition. To circumvent
this constraint, we propose a simple displacement on the unknown
function
\begin{equation}
 w (\theta) = U(\theta) + \delta,
\label{kb2}
\end{equation}
where $\delta$ is a constant to be determined. With this change, Eq.
(\ref{kb1}) becomes
\begin{equation}
\frac{d^2 U}{d \theta^2} - \frac{d U}{d \theta} + \left( (p-
2\delta)U - U^2 \right)  = 0, \label{kb3}
\end{equation}
where we have imposed
\begin{equation}
 k= p \delta - \delta^2.
\label{kb4}
\end{equation}
Now, we can factorize (\ref{kb3}) following the procedure exposed in
Section 2 with $F(U)=  (p- 2\delta)U - U^2$. Taking into account
Eqs. (\ref{n5})--(\ref{n6}), the functions $f_1(U)$ and $f_2(U)$
must be of the form
\begin{eqnarray}
&&f_1(U) = A\, U^{1/2} + B , \label{kb5}
\\
&& f_2(U)= (1-B)-\frac32 A\, U^{1/2}, \label{kb5bis}
\end{eqnarray}
with the following values of the two parameters $A$ and $B$
\begin{equation}
A^2=\frac23,\quad B=\frac25 \label{kb6}
\end{equation}
together with a constraint between the parameters $p$  and $\delta$,
introduced in Eqs. (\ref{L7}) and (\ref{kb2}), respectively:
\begin{equation}
 p = 2\delta + \frac6{25} \ .
\label{kb7}
\end{equation}
As we see from (\ref{kb7}), the coefficient $p$ (and therefore,
according to (\ref{L7}),  the velocity $v$ of the traveling wave)
depends on the parameter $\delta$, which is arbitrary since it is
fixed by  the integration constant $k$ in (\ref{kb4}).

Another important remark is that if we substitute this value of $p$
in (\ref{kb3}), we can realize that we get a kind of {\em universal
equation}, in which all the parameters have disappeared.

\begin{figure}[htp]
\centerline{
\includegraphics[scale=0.9]{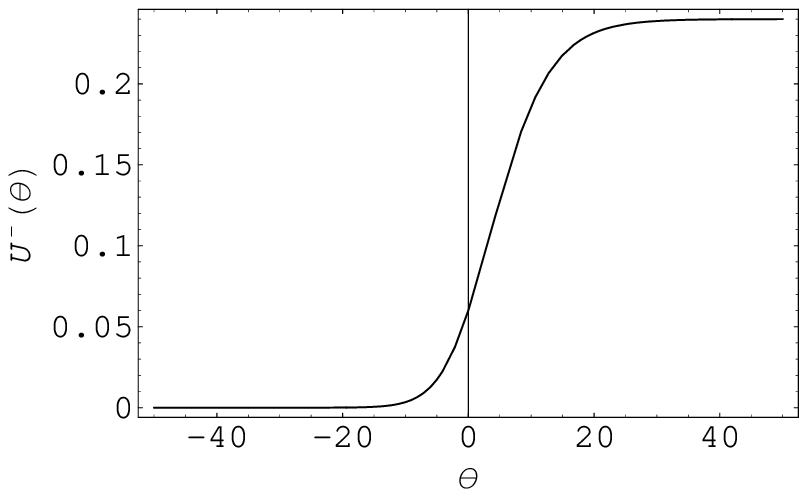}}
\caption{Plot of the {\em universal regular solution} $U^-(\theta)$
given in  (\ref{kb9}) with $\theta_0=0$.} \label{fig1}
\end{figure}

\begin{figure}[htp]
\centerline{
\includegraphics[scale=0.9]{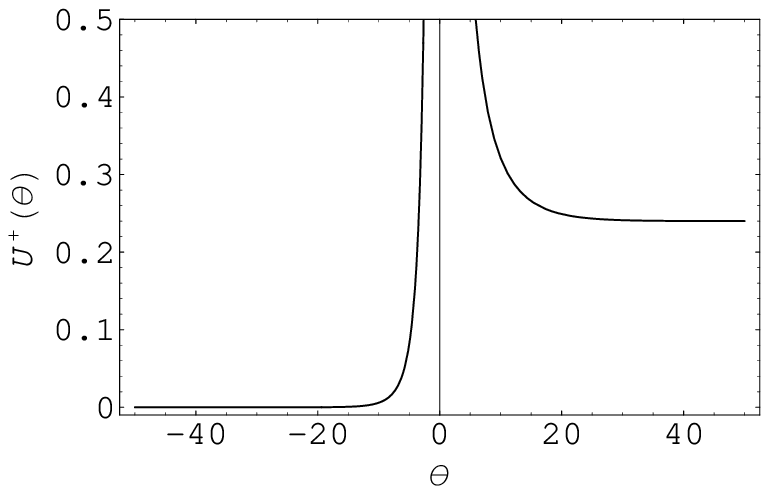}}
\caption{Plot of the {\em universal singular solution} $U^+(\theta)$
given in  (\ref{kb10}) with $\theta_0=0$.} \label{fig2}
\end{figure}

Particular solutions of (\ref{kb3}) are obtained by solving Eq.
(\ref{n7}) in this particular case, which turns out to be the
following couple of first order ODE of Bernoulli type:
\begin{equation}
 \frac{d U}{d \theta} = \pm \sqrt{\frac23}\ U^{3/2} + \frac25\, U ,
 \label{kb8}
\end{equation}
whose solutions, for the negative and positive sign, are
\begin{eqnarray}
&& U^-(\theta) = \frac{ 3}{50}\left[
 1+\tanh\left(\frac{\theta-\theta_0}{10}\right) \right]^2,
 \label{kb9}
 \\ [1ex]
&&U^+(\theta) = \frac{ 3}{50}\left[
 1+\coth\left(\frac{\theta-\theta_0}{10}\right) \right]^2 . \label{kb10}
\end{eqnarray}
In Figs.~\ref{fig1} and \ref{fig2} we show a plot of these  {\em
universal solutions}.

Returning to the original function $w(\theta)$ in (\ref{kb2}) is
trivial taking into account (\ref{kb7}). Then, using the definitions
introduced in Eqs. (\ref{L3}), (\ref{L5}) and (\ref{L7}), we get the
following family of regular solutions
\begin{equation}
u^-(x,t) = \frac{ v}{\alpha}   \label{kb26}
 + \frac{ 3\mu^2}{25 \alpha s}\left\{ \left[
 1+\tanh\left(\frac{\mu (x-vt -\xi_0)}{10 s} \right) \right]^2 -2 \right\}, 
\end{equation}
and the following family of singular solutions
\begin{eqnarray}
&&u^+(x,t) = \frac{ v}{\alpha}   \label{kb27} + \frac{ 3\mu^2}{25
\alpha s}\left\{ \left[
 1+\coth\left(\frac{\mu (x-vt -\xi_0)}{10 s} \right) \right]^2 -2 \right\}. 
\end{eqnarray}
Notice that after all these steps we have obtained a couple of
families (one regular and one singular) of particular solutions  of
the KdVB equation (\ref{L1}) depending on two free parameters, which
are the arbitrary constant $\xi_0$ and, what is more important, the
velocity $v$ of the travelling wave. In other words, our result
indicates that for every value of the velocity we have a family of
regular particular solutions of the KdVB equation (\ref{L1}), whose
particular features depend on the values of the parameters of the
original equation. However, in each case the solutions are always
kink-type solutions due to the presence of the  {\em universal
solution} $U^-(\theta)$ given in (\ref{kb9}), which is represented
in Fig.~\ref{fig1}.

It is interesting to compare our results with others obtained by
means of more elaborated or computerized methods. For instance, in
\cite{roychoudhury} the so-called {\em tanh-method} gives a number
of particular solutions called `regular', `singular' or `special'.
However, all these solutions are obtained in our result (\ref{kb26})
for particular choices of the parameters: their `regular' solution
(20) is obtained when we choose $\mu^2=100 s^2$, their `singular'
solutions correspond to the elections $\mu^2=25 s^2$ or $\mu^2=100
s^2$, and the `special' solution is found when the velocity of the
solitary wave is $v=6\mu^2/(25 s)$. Similarly, in
\cite{feng0,feng,chen}, using the `first integral method',
particular solutions are found and reported for specific values of
the velocity. From the standpoint of these methods, our solutions
(\ref{kb26}) and (\ref{kb27}), as well as other particular
solutions, can be obtained in the standard way that makes use of the
elementary symmetries of the KdVB equations such as shifts and
Galilean transformations. On the other hand, from the standpoint of
the factorization method the equivalent of these symmetries reduces
to attributing relative values to the parameters $\mu$ and $s$ as
stated above.

\begin{figure}[htp]
\centerline{
\includegraphics[scale=0.9]{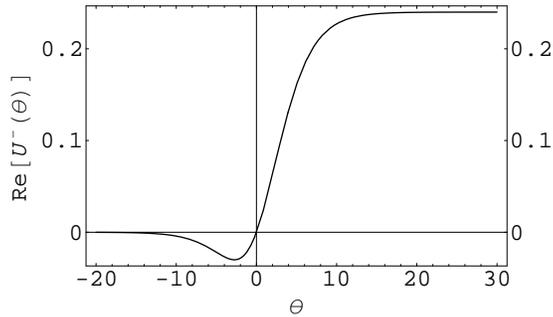}}
\caption{Plot of the real part of the {\em universal regular
solution} $U^-(\theta)$ given in  (\ref{kb9}) with
$\theta_0=-5i\pi/2$.} \label{fig3}
\end{figure}

\begin{figure}[htp]
\centerline{
\includegraphics[scale=0.9]{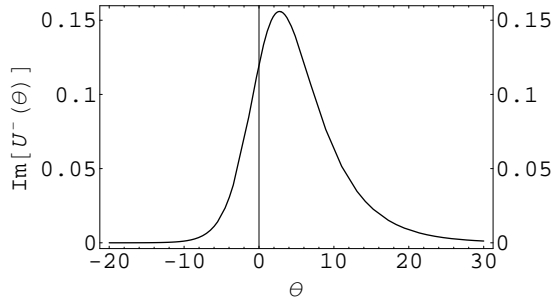}}
\caption{Plot of the imaginary part of the {\em universal regular
solution} $U^-(\theta)$ given in  (\ref{kb9}) with
$\theta_0=-5i\pi/2$.} \label{fig4}
\end{figure}

It is worth noting that if we consider the possibility of having
imaginary constants $\theta_0$ in (\ref{kb9})--(\ref{kb10}) or
$\xi_0$ in (\ref{kb26})--(\ref{kb27}), both types of solutions would
be in fact two versions of a unique solution. It is easy to show
that such imaginary constants are not forbidden mathematically by
the initial conditions. For example, the change $\theta_0\to
\theta_0+5i\pi $ in (\ref{kb9}) produces (\ref{kb10}). More
interesting situations are obtained for other complex values, for
instance if we choose $\theta_0=-5i\pi/2$ we get a complex solution
of the KdVB equation, whose real and imaginary parts are represented
in Figs.~\ref{fig3} and~\ref{fig4}, respectively. The behavior of
this complex solution, having the periodicity of the trigonometric
tangent introduced through the imaginary phase, is quite appealing
and is straightforwardly related with the complex scalar solutions
mentioned in Refs. \cite{liu,demiray}.

\begin{figure}[htp]
\centering{\begin{picture}(160,160)
\put(0,0){\includegraphics[height=60mm]{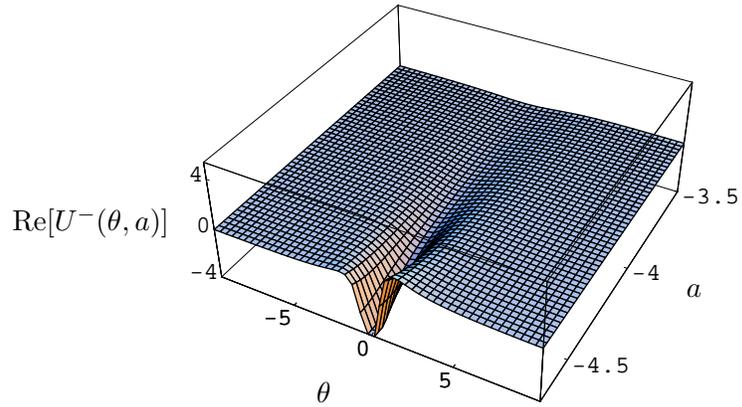}}
\put(50,10){$\theta$} \put(190,50){$a$} \put(-65,75){${\rm
Re}[U^-(\theta,a)]$}
\end{picture}
\caption{Plot of the real part of the {\em universal regular
solution} $U^-(\theta,a)$ given in  (\ref{kb9}).} \label{fig5} }
\end{figure}

\begin{figure}[htp]
\centering{\begin{picture}(160,160)
\put(0,0){\includegraphics[height=60mm]{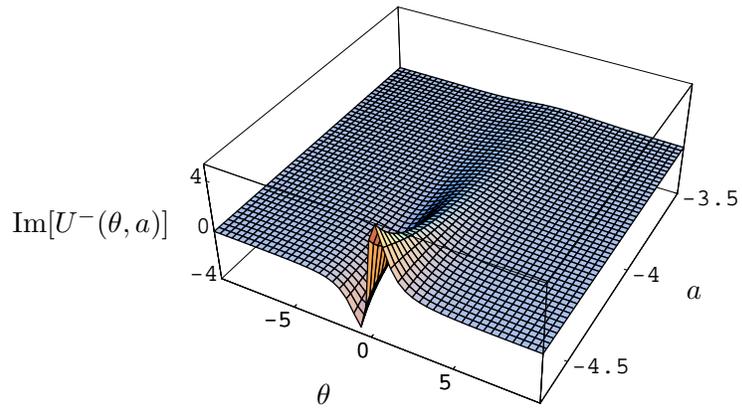}}
\put(50,10){$\theta$} \put(190,50){$a$} \put(-65,75){${\rm
Im}[U^-(\theta,a)]$}
\end{picture}
\caption{Plot of the imaginary part of the {\em universal regular
solution} $U^-(\theta,a)$ given in  (\ref{kb9}).} \label{fig6} }
\end{figure}

In fact, if we write $\theta _0=ia\pi$ (denoting now the solutions
as $U^-(\theta)\equiv U^-(\theta,a)$ to take into account the
initial condition), and examine the parametric dependence of the
real and imaginary part of the solution, we see that the real part
is passing from the pure kink type (for $a=0$) to the singular
solution for $a=-5$, whereas the imaginary part evolves from the
pure bell shape to more complex shapes at increasing $a$, see
Figs.~\ref{fig5} and \ref{fig6}, respectively.

Our three-dimensional plots are in agreement with Theorem 1 of Liu
and Liu \cite{liu} (see also their Remark 3) on complex KdVB
solutions that states that the real part is kink like whereas the
imaginary part is bell-shaped. However, we have found here that this
is a consequence of the transition of the regular solution towards
the singular one and vice versa when varying the amount of radians
in the imaginary phase. In addition, if one looks carefully to the
sections of constant $a$, one will find that there is a range of $a$
in which kink-type solutions with a pocket on the left tail and a
bump on the right tail occur. Such kinks have not been previously
reported in theory although it seems that they have been observed
experimentally \cite{nakamura}.

\section*{4. FACTORIZATION OF THE COMPOUND KdVB EQUATION}

 We will undertake the same program for the travelling-wave
solutions of the compound KdVB equation (\ref{L8}) with $q\neq0$. We
can factorize it directly as in Eq. (\ref{n2}), taking into account
that now $w=U$ and $F(U)=p\, U - U^2 -q\, U^3 - k$, by choosing the
factors in the form
\begin{eqnarray}
&&  f_{1}(U) \, U=A\,U^2+B\,U+C,\\ [1.ex] &&
f_{2}(U)=1-\frac{d(f_{1} U)}{dU} =-2A\,U +(1-B) . \label{c1}
\end{eqnarray}
The restriction conditions (\ref{n5})--(\ref{n6}) lead us to the
following values of the parameters:
 \begin{eqnarray}
&& A^2= \frac{q}2,\qquad B=\frac{A+1}{3A}, \label{c2a}
\\
&& C=\frac{1}{18} \left( \frac{2 - 9 p}A +\frac1{A^2} -
\frac{1}{A^3}\right) \label{c2}
\end{eqnarray}
together with
 \begin{equation}
\frac{1-2A}{3A} \, C= k. \label{c3}
\end{equation}
The last relation (\ref{c3}) means that the coefficient $C$ can take
any value, since $k$ is an arbitrary integration constant. This fact
means that we can get factorizations for any value of $p$, i.e., for
any velocity $v$ of the travelling wave.

Particular solutions of (\ref{L8}) are obtained by the corresponding
first order ODE (\ref{n7}), that in our case is
  \begin{equation}
\frac{d U}{d \theta} - A U^2 - B U - C=0 \label{c4}
\end{equation}
once the parameters $A,B,C$ are replaced by their values
(\ref{c2a})--(\ref{c2}). This is a simple Riccati equation whose
solutions are well known. One type of solutions is given by
 \begin{equation}
U^\pm(\theta) =\frac{-1}{3q} \pm \frac1{3\sqrt{2q}} \left[ 1+
\Delta\tanh\left( \frac{\Delta (\theta-\theta_0)}6 \right) \right]
\label{c5}
\end{equation}
where
 \begin{equation}
\Delta=\sqrt{18p+\frac{6}q -3}. \label{c6}
\end{equation}

\begin{figure}[htp]
\centerline{
\includegraphics[scale=0.9]{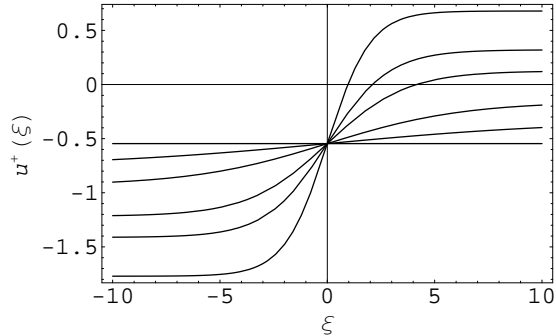}}
\caption{Plot of $u^+(x,t)\equiv u^+(\xi)$ given in (\ref{c7}) for
the following values of the parameters: $\alpha=3$, $\beta=2$,
$\mu=1$, $s=2$, $\xi_0=0$ and $v=-1.04,
-1.01,-0.94,-0.74,-0.54,-0.04$, being the constant solution the
corresponding to $v=-1.04$.} \label{fig7}
\end{figure}

Going back to the original variables of Eqs. (\ref{L3}), (\ref{L5})
and (\ref{L7}), the corresponding solutions of the original compound
KdVB equation (\ref{L2}) can be written as
 \begin{eqnarray}
&&u^\pm(x,t)  = -\frac{\alpha}{2\beta}  \label{c7}
\pm \frac{\mu}{\sqrt{6\beta s }} \left[ 1+ \Delta\tanh\left(
\frac{\mu\Delta (x{-}v t{-}\xi_0)}{6 s} \right) \right] ,
\end{eqnarray}
\begin{eqnarray}
\Delta  =\sqrt{\frac{18v s}{\mu^2}+\frac{9 s\alpha^2}{2\beta \mu^2}
-3}.
\end{eqnarray}
A plot of the functions $u^+(x,t)$ is given in Fig.~\ref{fig7} for a
certain choice of the parameters of the original KdVB equation and
different values of the wave velocity.

A second type of rational solutions of (\ref{c4}) appears when
$\Delta =0$, that is, for the particular value of $v$ given by
 \begin{equation}
6p = 1 -\frac2q \quad \text{or}\quad  v =\frac{\mu^2}{6 s} -\frac{
\alpha^2}{4\beta}. \label{xxx}
\end{equation}
This singular solution is
 \begin{equation}
U_0^\pm (\theta) = -\frac{k_0/A}{A + k_0\, \theta}-  \frac{A+1}{6
A^2}, \label{c39}
\end{equation}
where $A= \pm \sqrt{q/2}$ and $k_0$ is an integration constant. For
the special case where the free constant $k_0=0$, we get a constant
solution:
\begin{equation}
U^\pm_{0,0}(\theta) = - \frac{A+1}{6 A^2}~. \label{c8}
\end{equation}
Notice that this constant is directly obtained as a limit of the
solutions (\ref{c5}) when $\Delta \to 0$. If $k_0\neq 0$, the second
type solution (\ref{c8}) is recovered taking
\begin{equation}
\theta_0 = -\frac{A}{k_0} - \frac{3\pi i}{\Delta} \label{c9}
\end{equation}
and then the limit $\Delta \to  0$.

The rational solution (\ref{c39}) expressed in the initial variables
becomes
\begin{eqnarray}
u_0^\pm(x,t) = -\frac{\alpha}{2\beta}\left (  1\pm
\sqrt{\frac{2\beta\mu^2}{3s\alpha^2}}\right) \label{c10}
 - \frac{6 \alpha\mu k_0 }{2\beta\mu \pm k_0\sqrt{6s \beta\alpha^2} (x - v t-\xi_0)}~,
\end{eqnarray}
where $v$ is given in (\ref{xxx}) \footnote{If we introduce the
parameter $\epsilon=\mu\sqrt{\frac{2\beta}{3s\alpha ^2}}$ we can
write the rational solution in the simpler form $$ u_0^\pm(x,t)
=-\frac{\alpha}{2\beta}\Bigg[1\pm \epsilon +\frac{6\epsilon
k_0}{\epsilon \pm k_0(x-vt -\xi _0)}\Bigg]~,$$ where
$v=\left(\frac{\alpha}{2\beta}\right)^2[\epsilon ^2-1]$.}.

The first type solutions (\ref{c5}) were obtained in \cite{wang}
(with some mistakes, that have been corrected here) using the
`homogeneous balance method', and also in other papers
\cite{duffy,parkes,feng0,feng,chen} using different methods, but
after a considerable amount of computation. However, we have not
found any reference to the rational solutions.

\section*{5. FINAL REMARKS}

 In this paper, the factorization method has been shown to be well
adapted in the search of particular solutions of non-linear
differential equations of KdVB type. This method has a number of
advantages compared to others applied to the same problem: i) the
basic concept is quite simple and follows the same pattern already
used in linear equations; ii) the computations needed to develop it
are quite straightforward; iii) it allows to use analytical
arguments, for example when looking for the rational solutions of
the last section. Thus, the factorization technique is an
alternative method to other well-settled procedures and, as we have
shown here, can be successfully used to get exact particular KdVB
solutions. It can be applied to other non-linear partial
differential equations, looking in an efficient way for particular
solutions of physical interest. An immediate task is the
Kadomtsev-Petviashvili-Burgers (KPB) equation that is currently
considered to provide a more realistic framework for the study of
non-linear wave phenomena in the low and higher altitude auroral
region \cite{xue-kpb}.

\section*{ACKNOWLEDGEMENTS}
This work is supported by the Spanish M.E.C. (BFM2002-03773) and
Junta de Castilla y Le\'on (VA013C05) and partially supported by the
Mexican CONACyT project 46980. O.C.P. thanks the Department of
Theoretical Physics of the University of Valladolid for financial
support and kind hospitality.

\end{document}